\documentclass[epj]{svjour}
\usepackage{graphics}
\usepackage{graphicx}
\usepackage{color}
\definecolor{light-gray}{gray}{0.65}
\definecolor{webgreen}{rgb}{0,.5,0} 
\usepackage{dcolumn}
\usepackage{bm}
\usepackage{amssymb}
\begin{document}
\title{Evolution of Floquet Topological Quantum States\\ in Driven Semiconductors}
\author{Andreas Lubatsch\inst{1} \and Regine Frank\inst{2,}\inst{3,a}
}                     
\offprints{regine.frank@googlemail.com, regine.frank@rutgers.edu}          
\institute{Georg-Simon-Ohm University of Applied Sciences, Ke{\ss}lerplatz
  12, 90489 N\"urnberg, Germany \and Bell Laboratories, 600 Mountain Avenue, Murray Hill, NJ
07974-0636, USA \and Serin Physics Laboratory, Department of Physics and Astronomy,
Rutgers University, 136 Frelinghuysen Road, Piscataway, NJ 08854-8019, USA}
\date{Received: February 16, 2019 / Revised version: July 21, 2019}
%
\abstract{Spatially uniform excitations can induce Floquet topological
  bandstructures within insulators which have equal characteristics to those of
  topological insulators. Going beyond we demonstrate in this article
  the evolution of Floquet topological
  quantum states for electromagnetically driven semiconductor bulk matter. We show the direct
  physical impact of the mathematical precision of the Floquet-Keldysh
  theory when we solve the driven system of a generalized Hubbard model with
  our framework of dynamical mean field theory (DMFT) in the non-equilibrium. We explain the physical
  consequences of the topological non-equilibrium effects in our results for correlated sysems with impact on optoelectronic applications.}

\PACS{{71.10.-w}{Theories and models of many-electron systems}\and
{42.50.Hz}{Strong-field excitation of optical transitions in quantum systems; multi-photon processes; dynamic Stark shift}\and
{74.40+}{Fluctuations}\and
{03.75.Lm}{Tunneling, Josephson effect, Bose-Einstein condensates in periodic
potentials, solitons, vortices, and topological excitations}\and
{72.20.Ht}{High-field and nonlinear effects}\and {89.75.-k}{Complex systems}} 
\maketitle
\section{Introduction}
\label{INTRO}

Topological phases of matter \cite{Kosterlitz,KaneMele,HasanKane} have captured our imagination
over the past decades, revealing properties such as robust edge modes
and exotic non-Abelian excitations \cite{Fu,Moore}. Potential applications of
periodically driven quantum systems \cite{Dalibard} are
conceivable  from semiconductor spintronics \cite{Zuti} to topological quantum
computation \cite{Nayak}, topological lasers \cite{Rechtsman,Demetrios} and
random lasers \cite{Applsci2019} in
optics. The discovery of topological insulators in solid-state devices such as
HgTe/CdTe quantum wells \cite{Bernevig,Knig}, and topological Dirac insulators
such as Bi$_2$Te$_3$ and Bi$_2$Sn$_3$ \cite{Hsieh,Xia,Zhang} were milestones
in the research of the unique properties of topological phases in
technological applications.\\ 
Recently, it could be shown that the application of time-periodic
perturbations can induce topological properties in conventional insulators
\cite{Lindner1,Podolsky} which are trivial in equilibrium. So called Floquet topological
insulators include a wide range of physical solid state and atomic realizations,
driven both at resonance and off-resonance. These systems
display metallic conduction enabled by quasi-stationary states
at the edges \cite{Kitagawa,Gu,Lindner1}, Dirac cones in three dimensional
systems \cite{Gedik,Lindner2,Lindner3}, and Floquet Majorana fermions
\cite{Zoller}. Novel materials like Graphene are investigated as Floquet
fractional Chern insulators \cite{Rigol,Neupert,Bergholtz}.

\begin{figure}
\resizebox{0.48\textwidth}{!}{%
  \includegraphics{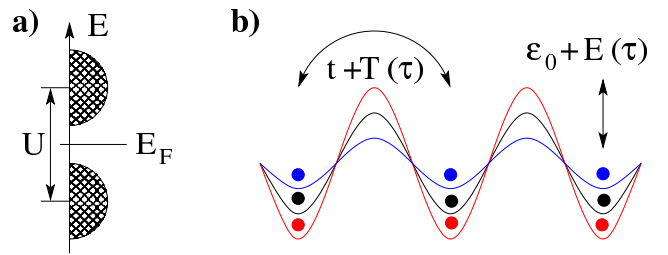}
}
\caption{Insulator to metal transition due to photo-excitation. 
(a) Schematic split in energy bands by the local Coulomb interaction $U$. The
width of the gap is determined symmetrically to the Fermi edge $E_F$. 
(b)  A periodic in time driving yields the
additional hopping contribution $T(\tau)$ of electrons in the lattice (black) and the local renormalization of the
lattice energy, the local potential, $E(\tau)$, as a quasi-energy. 
The colors in the lattice potential represent the external driving in time.}
\label{Mott}       
\end{figure}

In this article we show that Floquet topological quantum states evolve in
correlated electronic systems of driven conventional semi-conductors at low temperature. The non-equilibrium is
in this sense defined by the intense external driving field which
induces topologically dressed electronic states as well as dynamical gaps.  We
show that the mathematical method of the expansion into Floquet modes
\cite{Floquet} is leading to results of direct physical impact, when we
implement it for modeling the coupling of a classical electromagnetic external
driving field to the correlated quantum many body system. Our results derived by
Dynamical Mean Field Theory (DMFT) in the non-equilibrium provide novel
insights in topologically induced phase transitions of driven and otherwise
conventional 3-dimensional semiconductor bulk matter and insulators.

\begin{figure}
\resizebox{0.45\textwidth}{!}{%
  \includegraphics{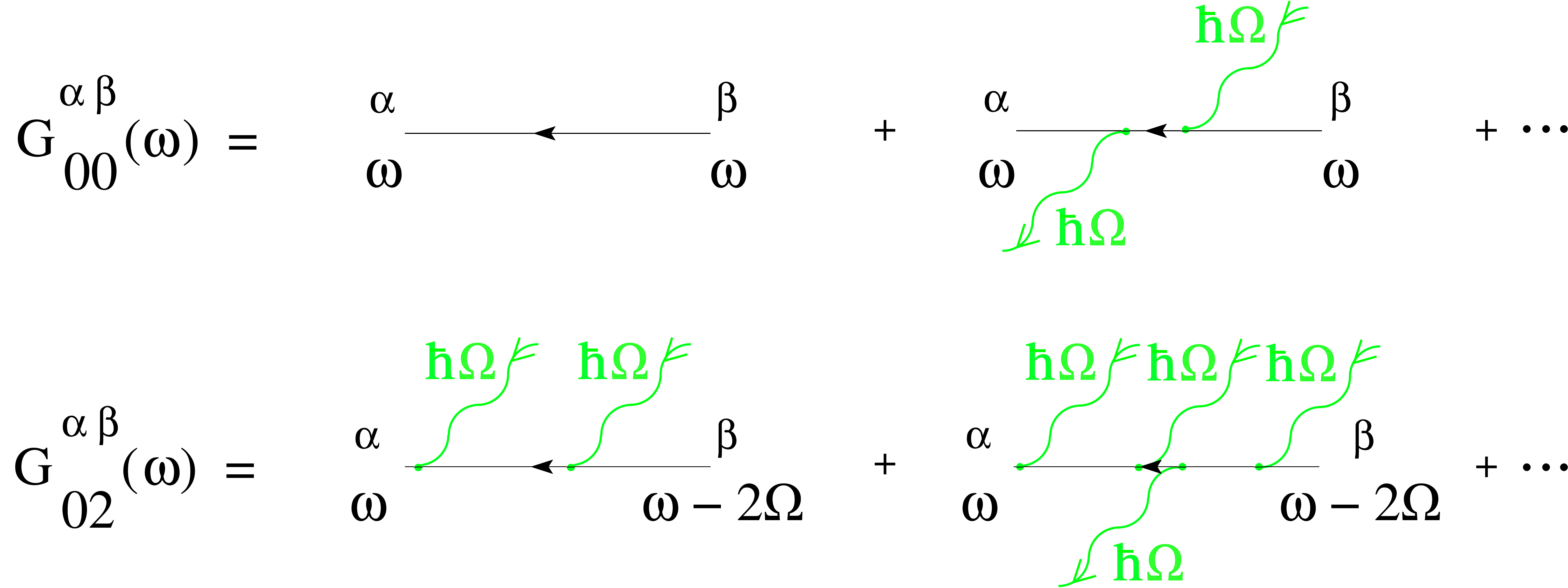}}
\vspace{0.5cm}       
\caption{The contributions to the Floquet Green’s function in terms of absorption and emission of
  external energy quanta $\hbar\Omega_L$ are schematically explained. $G^{\alpha\, \beta}_{00} (\omega)$ represents the sum of all balanced
  processes, $G^{\alpha\, \beta}_{02} (\omega)$ describes a net absorption of
  two photons.  $\alpha, \beta$ are Keldysh indices.}
\label{FloquetGreen}       
\end{figure}

\section{Quantum Many Body Theory for Floquet Topological States in the Non-Equilibrium}
\label{sec:HUBBARDMODEL}

\begin{figure}
\resizebox{0.5\textwidth}{!}{\includegraphics{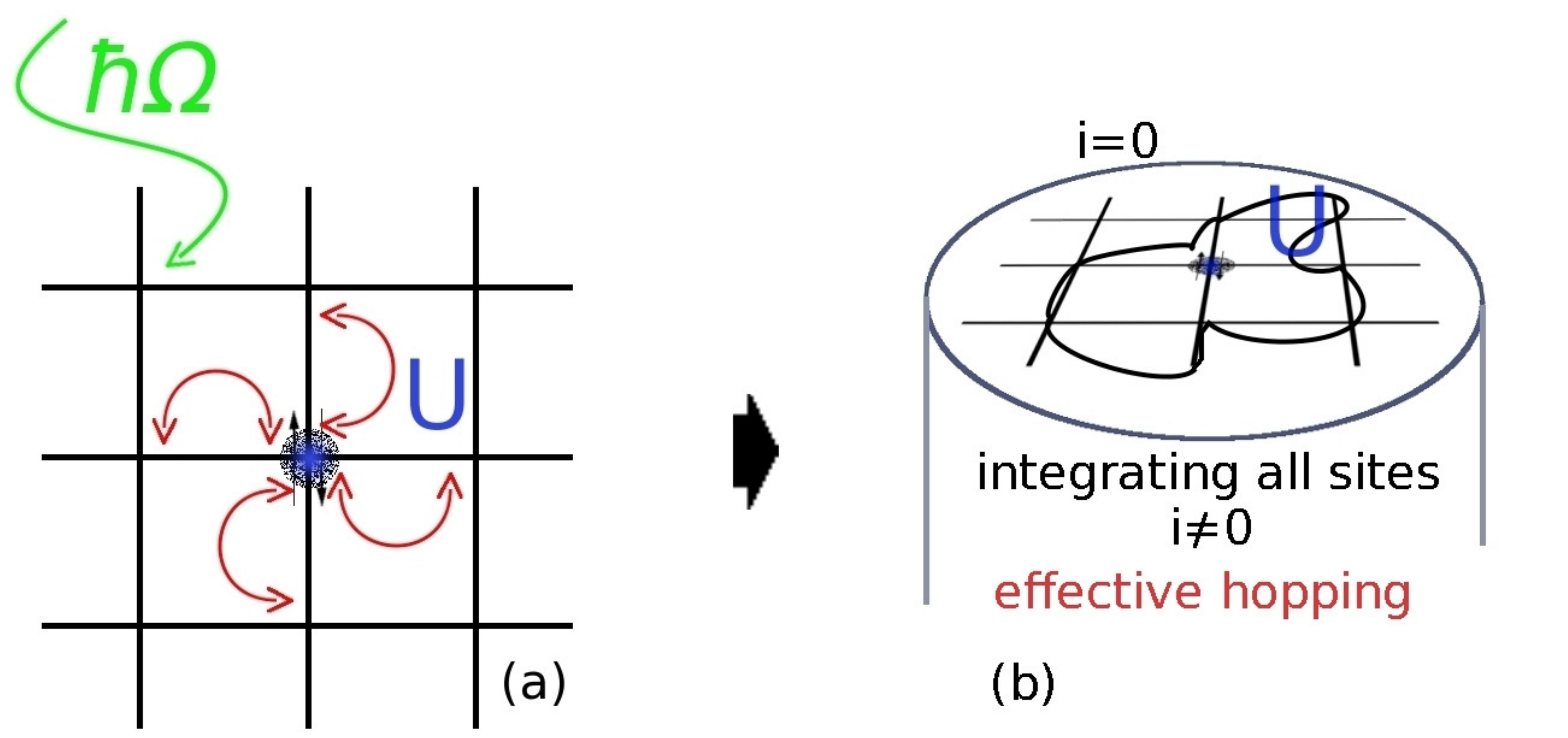}}
\vspace{0.5cm}       
\caption{Schematic representation of the non-equilibrium dynamical mean field theory. 
(a) The semiconductor behaves in the considered regime as an insulator: In the
non-equilibrium DMFT-scheme optical excitations by external photons of energy $\hbar \Omega$
yield additional electronic hopping processes. 
(b) DMFT idea: Integration over all lattice sites leads to
an effective theory including non-equilibrium excitation. The bath consists of
all single sites and the approach is thus self-consistent. 
The driven electronic system may in principal couple to a surface-resonance or
an edge state. The coupling to these states can be enhanced by the external excitation
of the electronic many body system.}
\label{DMFT}       
\end{figure}

We consider in this work the wide gap semiconductor bulk as it is driven by a
strong periodic in time external field in the optical regime which yields
higher-order photon absorption processes. The electronic dynamics of the
photo-excitation processes, compare with Fig. \ref{Mott}, is modeled as a
generalized driven Hubbard Hamiltonian Eq. (\ref{Hamilton_we}). The system is solved by applying a
  Keldysh formalism including the electron-photon interaction in the sense of
  coupling the electromagnetic field to the electronic dipole and thus to the
  electronic hopping as an additional kinetic contribution, compare figure \ref{FloquetGreen}. We implement a
  dynamical mean field theory (DMFT), compare with Fig. \ref{DMFT}, with a generalized iterative perturbation theory solver (IPT), see
  Fig. \ref{IPT-Sigma}, in order to solve the system numerically which will
  be explained in detail. We start with the introduction of the full interacting Hamiltonian, Eq. (\ref{Hamilton_we}), 
\begin{eqnarray}
\!\!\!\!\!\!\!\!H\!&=&\!\! \sum_{i, \sigma} \! \varepsilon_i  c^{\dagger}_{i,\sigma}c^{{\color{white}\dagger}}_{i,\sigma} 
+   \frac{U}{2} \sum_{i, \sigma} c^{\dagger}_{i,\sigma}c_{i,\sigma}c^{\dagger}_{i,-\sigma}c_{i,-\sigma}\label{Hamilton_we}
\\&& - t\!\! \sum_{\langle ij \rangle, \sigma}\!\!
c^{\dagger}_{i,\sigma}c^{{\color{white}\dagger}}_{j,\sigma}\nonumber
\\&& + i\vec{d}\cdot\vec{E}_0 \cos(\Omega_L \tau)\sum_{<ij>,\sigma }
 \left(
           c^{\dagger}_{i,\sigma}c^{{\color{white}\dagger}}_{j,\sigma} 
 	  -
           c^{\dagger}_{j,\sigma}c^{{\color{white}\dagger}}_{i,\sigma} \right).\nonumber
\end{eqnarray}
In Eq. (\ref{Hamilton_we})  $c^\dagger,(c)$ denote the creator (annihilator) of an electron. The subscripts $i,j$ indicate the site, $\langle i,j \rangle$ implies the sum over
nearest neighboring sites. The term $\frac{U}{2} \sum_{i,
  \sigma}c^{\dagger}_{i,\sigma}c_{i,\sigma}c^{\dagger}_{i,-\sigma}c_{i,-\sigma}$
is devoted to repulsion due to the onsite Coulomb interaction $U$ between electrons with opposite spins. The third
term $-t \sum_{\langle ij \rangle, \sigma}\!\!
c^{\dagger}_{i,\sigma}c^{{\color{white}\dagger}}_{j,\sigma}$ describes a
standard hopping of electrons with the amplitude $t$ between nearest neighbor sites. Those
contributions form the standard Hubbard model which we generalize for our
purposes in what follows. The first term $\sum_{i, \sigma} \! \varepsilon_i
c^{\dagger}_{i,\sigma}c^{{\color{white}\dagger}}_{i,\sigma}$   generalizes the
Hubbard model with respect to the onsite energy. The electronic
on-site energy is noted as $\varepsilon_i$. 
The central feature, the external time-dependent electromagnetic driving is described in terms
of the field $\vec{E}_0$ with laser frequency $\Omega_L $, $\tau$. We model the photonic interaction as a
coupling of the amplitude of the impinging electromagnetic wave and the
electronic dipole operator $\hat{d}$ with strength $|\bf{d}|$.
The expression $ i\vec{d}\cdot\vec{E}_0 \cos(\Omega_L\tau)\sum_{<ij>,\sigma} 
 \left(c^{\dagger}_{i,\sigma}c^{{\color{white}\dagger}}_{j,\sigma} -
   c^{\dagger}_{j,\sigma}c^{{\color{white}\dagger}}_{i,\sigma}\right)$ describes the renormalization of the standard electronic
         hopping processes, as a possible contribution $T(\tau)$ in Fig. (\ref{Mott}),
         due to interaction with the external pump field.

\subsection{COUPLING Of A CLASSICAL DRIVING FIELD TO A QUANTUM DYNAMICAL SYSTEM}
\label{sec:FLOTHEORY}

In order to solve the generalized Hubbard Hamiltonian Eq. (\ref{Hamilton_we}) for the driven
system, we briefly introduce the explicit time
dependency of the external field. It yields Green's
functions which depend on two separate time arguments. 

The double Fourier transform from time to frequency coordinates leads to two
frequencies which are chosen as the relative and the center-of-mass frequency \cite{PRB,FrankANN} and we
introduce an expansion into Floquet modes
\begin{eqnarray}
\label{Floquet-Fourier}
G_{mn}^{\alpha\beta} (\omega) &=&
\left\lmoustache \!\!{\rm d}{\tau_1^\alpha}\!{\rm d}{\tau_2^\beta}\right.
e^{-i\Omega_L(m{\tau_1^\alpha}-n{\tau_2^\beta})}
e^{i\omega({\tau_1^\alpha}-{\tau_2^\beta})}
G (\tau_1^\alpha,\tau_2^\beta)\nonumber\\
&\equiv&
G^{\alpha\beta} (\omega-m\Omega_L, \omega - n\Omega_L).
\end{eqnarray}
Generally Floquet \cite{Floquet} states are analogues to Bloch states. Whereas the Bloch
state results from a spatially periodic potential the Floquet state is a result of
the temporal periodicity \cite{Faisal,Grifoni,Brandes,Eckardt,Uhrig,Sentef,Fan,PRB,FrankANN,ANN}. Either induces a
specific topological band-structure. The Floquet expansion is introduced
in Fig. \ref{FloquetGreen} as an intelligible graphical explanation of what is
described in Eq. (\ref{Floquet-Fourier}). $(m,n)$ label the
Floquet modes, whereas $(\alpha,
\beta)$ refer to the branch of the Keldysh contour ($\pm$) where the
respective time argument resides. The result of this process, the Floquet
modes or Floquet quasi energies are derived by a
Fourier transformation of the periodic in time potential as the principal
structure of bands in frequency space. It is basically the analogue
procedure in time to the calculation of the band structure due to periodic in
space potentials in solid state materials. The physical consequence of the process
however is noteworthy. It can be understood as the quantized absorption and
emission of energy
$\hbar \Omega_L$ out of and into the classical external driving field by the
driven quantum many body system.

In the analytical case of uncorrelated electrons, $U\,=\,0$, the solution for the Green's
function  $G_{mn}(k,\omega) $ can be found directly by solving the Hamiltonian which eventually reads for the retarded component
\begin{eqnarray}
G_{mn}^{R}(k,\omega) \label{Green}
=
\sum_{\rho}
\frac
{
J_{\rho-m}\left(A_0\tilde{\epsilon}_k \right)
J_{\rho-n}\left(A_0\tilde{\epsilon}_k \right)
}
{
\omega -\rho\Omega_L - \epsilon_k + i 0^+
}.
\end{eqnarray}
Here $\tilde{\epsilon}_k$ represents the dispersion relation which is induced
by the external driving field and which is to be distinguished from the
lattice dispersion $\epsilon_k$. $J_n$ are the cylindrical Bessel functions of integer order, $A_0 = \vec{d}\cdot\vec{E}_0 $ and
$\Omega_L$ is the laser frequency as introduced before.
The retarded Green's function for the optically excited band electron is then given by

\begin{eqnarray}
\label{EqGsum}
G^{R}_{\rm Lb}(k,\omega)  
=
\sum_{m,n}
G_{mn}^{R}(k,\omega).
\end{eqnarray}

\subsection{NUMERICAL SOLUTIONS: DYNAMICAL MEAN FIELD THEORY IN THE NON-EQUILIBRIUM}
\label{sec:NUMSOL}

\begin{figure}
\hspace*{0.0cm}\resizebox{0.48\textwidth}{!}{%
  \includegraphics{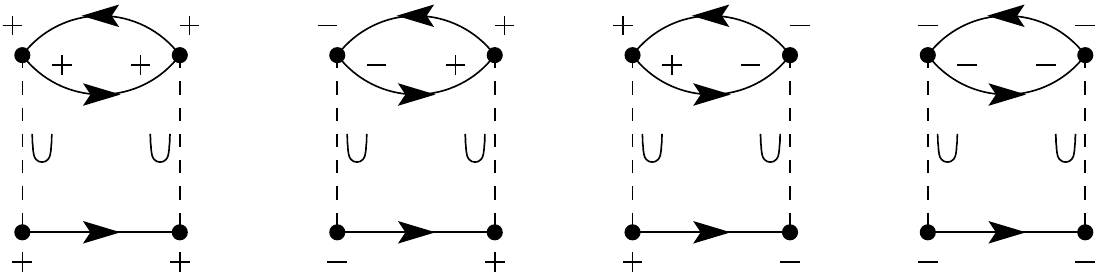}
}
\vspace{0.5cm}       
\caption{Local self-energy $\Sigma^{\alpha \beta}$ within the iterated
  perturbation theory (IPT). The used IPT is a second order
  diagrammatic solver with respect to the electron electron interaction $U$
  and it is here 
generalized to non-equilibrium, $\pm$ signs are indicating the respective branch of the Keldysh contour. 
The solid lines represent the bath-Green's function, the so called Weiss-field
$\mathcal{G}^{\alpha\beta}$, see ref. \cite{NJP}.}
\label{IPT-Sigma}       
\end{figure}

The generalized Hubbard model for the correlated system at hand, $U\,\neq\,0$,
in the non-equilibrium, Eq. (\ref{Hamilton_we}), is numerically solved by
means of a single-site Dynamical Mean Field Theory (DMFT)
\cite{ANN,Kotliar,Metzner,Monien,Hettler,Freericks,OKA,Reichmann,Zgid,Werner,APLB,NJP,Max,Walter1,Walter}. As explained in section \ref{sec:FLOTHEORY}  the expansion
into Floquet modes in combination with the proper Keldysh description
implements the external time dependent classical driving field, and couples it to the
quantum many body system. The Floquet-Keldysh DMFT 
\cite{ANN,NJP} is solved with a second order iterative perturbation theory (IPT), the
contributing diagrams to the local self-energy $\Sigma^{\alpha\beta}$ are
explained in Fig. \ref{IPT-Sigma}. 
The Green's function $G^{R}_{\rm Lb}(k,\omega)$ for the interaction of the
band electron with the external laser, Eq. (\ref{EqGsum}), is 
characterized by the wave vector $k$, which describes the periodicity of the
lattice, and it depends on the electronic frequency $\omega$ and
the external driving frequency $\Omega_L$, see Eq. \ref{Floquet-Fourier}, that
are captured in the Floquet indices $(m,n)$. For the requirements of our system
the DMFT self-consistency relation, the actual centerpiece of the DMFT scheme,
assumes the form of a matrix equation of the non-equilibrium Green's
functions which is of dimension $2 \times 2$ in regular Keldysh space and of
dimension $n \times n$ in Floquet space.  The resulting numerical algorithm
proves to be efficient and stable also for all values of the Coulomb interaction $U$.

Whereas in previous work \cite{ANN,NJP,APLB} we considered an additional kinetic energy contribution due to a lattice vibration, here we take into account  a
coupling of the microscopic electronic dipole moment to an external
electromagnetic field \cite{PRB,FrankANN} for the correlated system. This requires the introduction
of the quantum-mechanical expression for the electronic dipole operator  $\hat d$, see
the last term r.h.s. Eq.(\ref{Hamilton_we}), $i\vec{d}\cdot\vec{E}_0 \cos(\Omega_L \tau)\sum_{<ij>,\sigma}
 \left(
           c^{\dagger}_{i,\sigma}c^{{\color{white}\dagger}}_{j,\sigma} 
 	  -
           c^{\dagger}_{j,\sigma}c^{{\color{white}\dagger}}_{i,\sigma}\right)$. This contribution is conceptionally different from the generic kinetic
hopping of the third term of Eq.(\ref{Hamilton_we}). The coupling $\hat d
\cdot \vec E_0\cos(\Omega_L\tau)$  generates the factor
$\Omega_L$ that cancels the $1/\Omega_L$ in the renormalized cylindrical
Bessel function in Eq. (7) of ref. \cite{NJP} under the assumption of the
Coulomb gauge, 
$\vec E(\tau)\,= - \frac{\partial}{\partial \tau} \vec A(\tau)$, which reads
in Fourier space as $\vec E(\Omega_L)\,=\,i\Omega_L\cdot\,\vec A(\Omega_L)$. The consistency check of the Floquet sum is discussed in section \ref{sec:FLO}
here in detail.
\begin{figure}
\resizebox{0.5\textwidth}{!}{\includegraphics{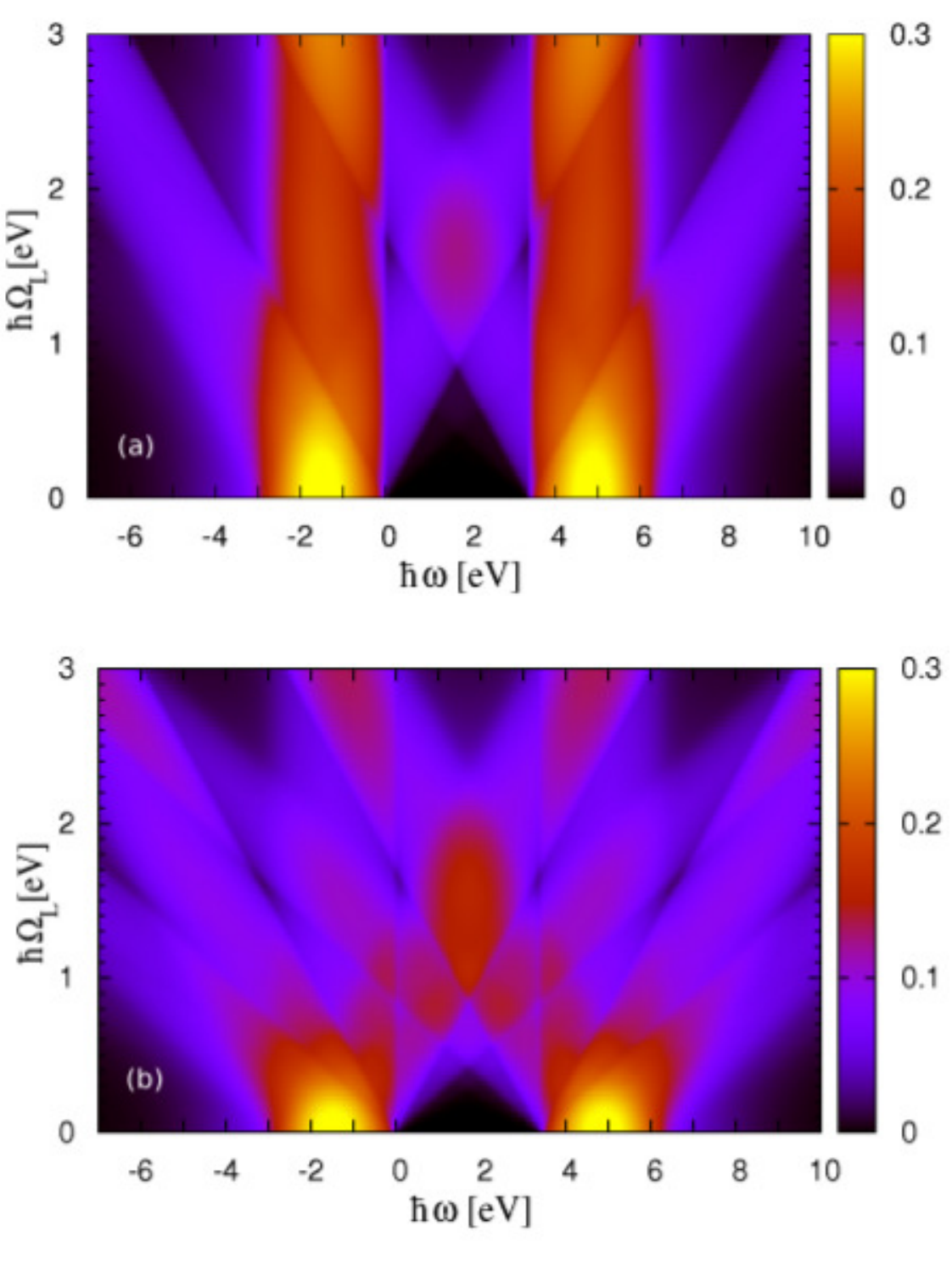}}
\vspace*{0.5cm}     
\caption{Energy spectra of Floquet topological quantum states of the
  semiconductor bulk in the non-equilibrium. (a) The evolution of the LDOS in the non-equilibrium for varying
  excitation laser frequencies $\Omega_L$ is displayed. The excitation
  intensity of 5.0 $MW/cm^{2}$ is kept constant. The original bandgap in the
  equilibrium, see Fig. \ref{Mott}, is vanishing with rising driving frequency
  and dressed states emerge due to the non-equilibrium AC-Stark effect \cite{SchmittRink,Chemla}. The
  split bandstructure is replaced by a Floquet fan which evolves with
  increasing external excitation frequency. The intersection of bands is observed
  and the formation of topological subgaps, see e.g. at $\hbar\Omega_L\,=\,1.75\,eV$
  can be observed.  (b) The evolution of the LDOS in the non-equilibrium for varying
  excitation laser frequencies $\Omega_L$ is displayed, the excitation
  intensity of 10.0 $MW/cm^{2}$ is assumed. Spectral weight is shifted to a
  multitude of Floquet-bands which clearly emerge, while the original split
  band characteristics almost vanishes apart from the near-gap band edges. The
Floquet bands overlap with each other and a variety of Floquet gaps are
formed. Multiple induced band edges are found and their crossing points can be
identified, where topologically induced transitions are possible, especially the
generation of higher harmonics can be enhanced at such points.
  For a detailed discussion see section \ref{sec:QUASIENERGY}.}
\label{IMEffectiveGFD}       
\end{figure}

\begin{figure}
{\hspace*{-1.0cm}\resizebox{0.55\textwidth}{!}{\includegraphics{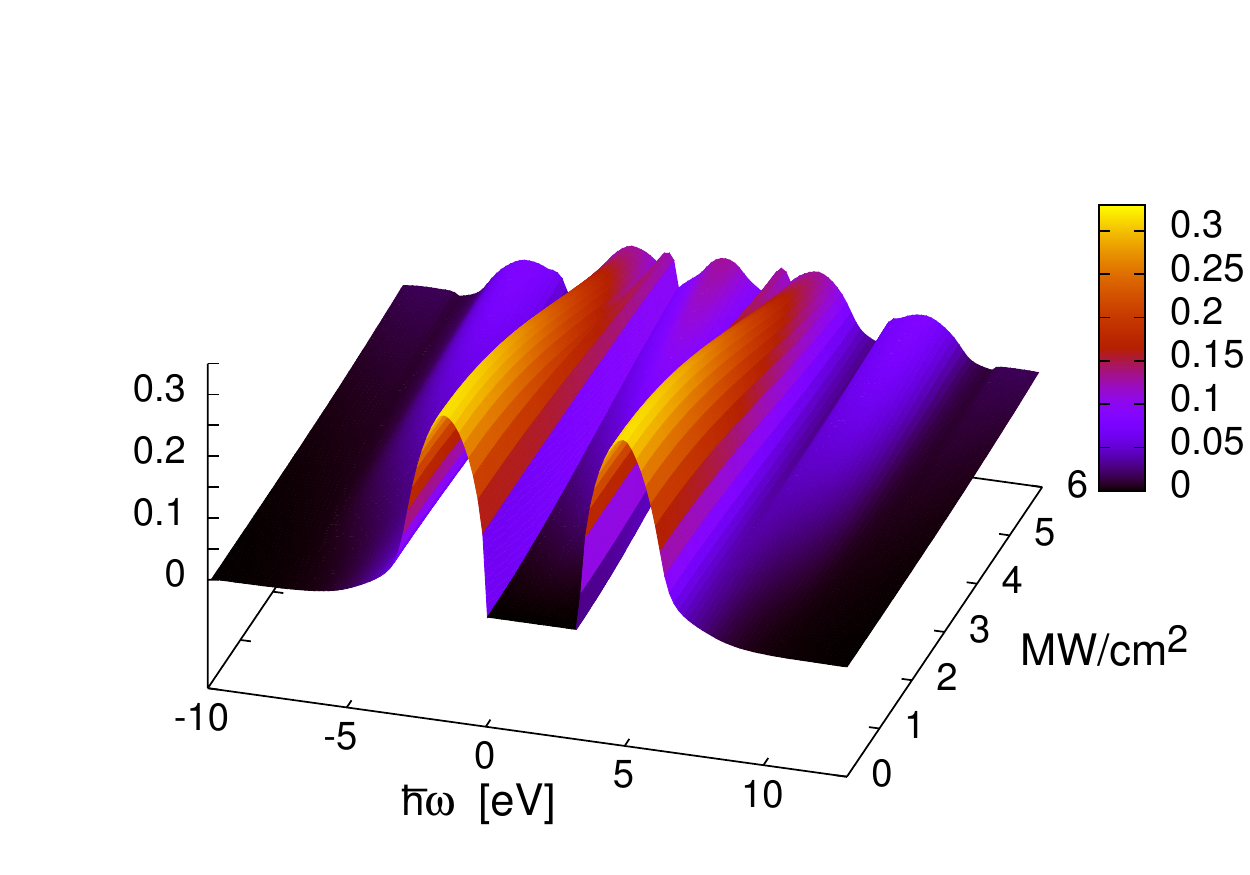}}}
\caption{Energy spectra of Floquet topological quantum states of the
  semiconductor bulk in the non-equilibrium. The evolution of the LDOS is
  displayed for the single excitation energy of $\hbar\Omega_L=1.75 eV$, wavelength
  $\lambda=710.0 nm$ and an increasing external driving intensity up to $6.0
  MW/cm^2$. Spectral weight is shifted by excitation to emerging Floquet
  sidebands and a sophisticated sub gab structure is developed. In the non-equilibrium such
  topological effects in correlated systems are non-trivial. The
  bandgap in equilibrium is $3.38 eV$, the Fermi edge is $1.69 eV$, the width
  of each band is $3.0 eV$. For a detailed discussion see section \ref{sec:QUASIENERGY}.}
\label{IMEffectiveGAD}       
\end{figure}
It has been shown by \cite{Monien} that the
  coupling of an electromagnetic field modulation to the onsite electronic density $n_i\,=\,c^{\dagger}_{i,\sigma}c_{i,\sigma}$ alone as an effect in the unlimited three
  dimensional translationally invariant system, can be gauged away. The effect
  for this
  type of coupling can be absorbed in an overall shift of the local potential
  while no additional dispersion is reflecting any induced functional dynamics of the system. Thus
  such a system \cite{Rigol,Bukov} is obviously not
  going to show any topological effects in the sense of a topological
  insulator or a Chern insulator. By contrast, the coupling of the external
  electromagnetic field modulation to the dipole moment of the
  elementary charge, and thus to the hopping, see Eq. (\ref{Hamilton_we}), in the sense of a kinetic energy of the fermions, cannot be gauged away and it
  is causing the evolution of topological states in the three dimensional
  unlimited system. Line three of Eq. (\ref{Hamilton_we}) formally represents
  the electromagnetically induced kinetic contribution
\begin{eqnarray}
i\vec{d}\cdot\vec{E}_0\cos(\Omega_L\tau)\sum_{<ij>,\sigma}\left(c^{\dagger}_{i,\sigma}c^{{\color{white}\dagger}}_{j,\sigma}
  - c^{\dagger}_{j,\sigma}c^{{\color{white}\dagger}}_{i,\sigma}\right)\,=\,\\
e\sum_{\vec{r}}\hat{j}_{ind}(\vec{r})\cdot\vec{A(\vec{r},\tau)}\nonumber
\label{curcont}
\end{eqnarray}

which is equivalent to a kinetic contribution of the photo-induced charge
current dependent in space with $\vec r$
\begin{eqnarray}
\vec{j}_{ind}(\vec{r})_{\delta}\,=\,-\,\frac{t}{i}\sum_{\sigma}(c^{\dagger}_{\vec
  r, \sigma}c_{\vec r +\delta, \sigma}\,-\,c^{\dagger}_{\vec r +\delta,
\sigma}c_{\vec r, \sigma}).
\end{eqnarray}
The temporal modulation of the
         classical external electrical
         field in (111) direction always comes along with a temporally
         modulated magnetic field contribution $\vec B (\vec
         r,\tau)\,=\,\nabla\times\vec{A}(\vec r,\tau)$ with $\vec B
         (\vec{r},\Omega_L)$ in Fourier space, as a
         simple consequence of Maxwells equations. In what follows we derive
         the non-equilibrium local density of states (LDOS) as the central
         result of DMFT which comes along with the dynamically evolving
         life-time of non-equilibrium states as the inverse of the imaginary part of the
         self-energy $\tau\,\sim\,1/\Im\Sigma^R$. A time reversal
         procedure derived by an external field will in any case not be able
         to revise the non-equilibrium effect. The light-electron
         coupling will be modified and the overall profoundly differing
         material characteristics, the conductivity and the polarization
         of the excited matter in the non-equilibrium are preventing any
         time-reversal process in the sense of closing the Floquet fan again in this regime. The initially impinging
         electromagnetic field thus causes
         a break of the time-reversal symmetry, the current ultimately leads to the acquisition of a
         non-zero Berry flux. A Wannier-Stark type ladder
         \cite{Zak89} is created that can be characterized by the Berry phase
         \cite{BerryElectronicProperties} in the sense of a Chern or a winding
         number and the $\mathbb{Z}_2$ invariants in three dimensions
         respectively \cite{Vanderbilt2017}.

\section{Floquet Spectra for Driven Semiconductors}

\label{sec:QUASIENERGY}

From the numerically computed components of the Green's function, 
we define the local density of states $N(\omega,\Omega_L)$ where  momentum is integrated out and Floquet indices are summed
\begin{eqnarray}
N(\omega,\Omega_L)
=
-\frac{1}{\pi}\sum_{mn} \int {\rm d^3}k  {\rm Im\,} 
G^{R}_{mn} ({\bf k},\omega, \Omega_L).
\label{Eq:DoS}
\end{eqnarray}

In the following we show results for optically excited semiconductor bulk
matter, with a band gap in the equilibrium of $3.38\,eV$, typical parameters
for ZnO. ZnO \cite{JiaLu1,JiaLu2,ZnOMono1,ZnOMono2} is a very promising material for the construction of
micro-lasers, quantum wells and optical components as well as for the
engineering of ultrafast switches in certain geometries and in
connection to other
topological insulators. It is broadly investigated in the  non-centro-symmetric wurtzite configuration as well as in
the centro-symmetric, cubic, rocksalt configuration
\cite{Rocksalt1,Rocksalt2}, it's band-\\gap is estimated to be of 2.4 eV up to 6.1 eV depending on various factors
as the pressure during the fabrication process. Either crystal configuration is a good candidate for the production of
second or higher
order harmonics under intense external excitations \cite{Wegener}. Thus it is
of high interest for this work.

\subsection{Evolution of Floquet Topological Quantum States in the
  Non-Equilibrium}
\label{sec:FLO}

In Fig. \ref{IMEffectiveGFD} we investigate a wide gap semi-conductor structure,
the band gap in equilibrium is assumed to be $3.38\,eV$, which shall be exposed to an
external periodic-in-time driving field. The system is so far
considered as pure bulk, so we are investigating pure Floquet topological
effects in the non-equilibrium. The excitation intensity is
considered to be $5.0 MW/cm^2$ in the results of Fig. \ref{IMEffectiveGFD}(a)  and $10.0
MW/cm^2$ in the results of Fig. \ref{IMEffectiveGFD}(b). DMFT as
  a solver for correlated and strongly correlated electronics as such is a
spatially independent method and thus it is designed to derive bulk effects,
whereas all k-dependencies have been
integrated as a fundamental aspect of the methodology. Thus we are not analyzing the k-resolved information of the
Brillouin zone, DMFT results in one, two and three dimensions are independent of
any spatial information as long as no artificial coarse graining with a novel
length scale in the sense of finite elements or finite volumes is included. The sheer fact however, that the energy dependent LDOS profoundly
changes with a varying excitation frequency and with a varying excitation intensity
as well, proves that also the underlying k-dependent bandstructure is
topologically modulated and a non-trivial topological structure of the Hilbert
space is generated by external excitation. Our system in equilibrium is fully
periodic in space and time. The dependent-in-time  external electrical field
generates an temporally modulated magnetic field and thus a dynamical
Wannier-Stark effect. Floquet states are generated. They are the temporal
analogue to Bloch states, and the argumentation by Zak
\cite{Zak89} applies for the generation of the Berry phase $\gamma_m$ when a
solid is exposed to an externally modulated electromagnetic potential
\cite{BerryElectronicProperties,Vanderbilt1,Vanderbilt2,Resta}. The Floquet quasi-energies, the topology, see
Fig. \ref{FloquetGreen}, are labeled by the Floquet modes in dependency to the external excitation frequency,
and to the external excitation amplitude. The topological invariants, the
Chern number as a sum over all
occupied bands $n\,=\,\sum_{m=1}^{\nu}n_m \neq 0$ and the $\mathbb{Z}_2$
invariants include the Berry flux $n_m\,=\,1/2\pi\int d^2 \vec{k} (\nabla
\times \gamma_m)$. The winding number is also associated consistently with the argument of
collecting a non-zero Berry flux. The driving frequency is
considered to be smaller than the width of the semiconductor gap in
equilibrium. While the system is excited and thus evolving in
  non-equilibrium, a Berry phase is acquired and a non-zero Berry flux und
  thus a non-zero Chern number are characterizing the topological bandstructure as {\it
    non-trivial}. 
For
  one dimensional models \cite{Zak89,DalLago} it is well acknowledged that
  with the reduction of the external excitation frequency $\Omega_L$ novel
  sets of Floquet bands, so called replica, emerge in the spectrum, which are
  attributed a quantised change of the Berry phase $\gamma\,=\,\pi$. In three
  dimensions the situation as such is more complicated,
  the Berry phase is
  usually attributed to the Wyckoff positions of the crystal and the Brioullin
  zone respectively \cite{Zak89}, and as such it cannot be derived by
  the pure form of the DMFT. $\vec k$-dependent information can be derived
  by so called real-space or cluster DMFT solutions (R-DMFT or CDMFT)
  \cite{Gull,Hofstetter,Peters1,Peters2}, however up to our best
  knowledge a powerful solution for true three-dimensional systems in the
  non-equilibrium is not existing in presence. 
These thoughts however do not stand against the logical consequence that a system out of
  equilibrium acquires a non-zero Berry phase. The
  inclusion of 
  interactions leads to a Mott-type gap which closes due to the superposition
  of the crossing Floquet bands, however also the opening of non-equilibrium induced Mott-gap
  replica can be found for $\Omega_L\,=\,1.69\,eV$, which is half of the width
of the Mott-gap. The replica are complete at $\Omega_L\,=\,3.38\,eV$, see
Fig. \ref{IMEffectiveGFD}(a). For the increase of $\vec E_0$ these gap replica are
again crossed by the next order of Floquet sidebands, however their width at
$\Omega_L\,=\,3.38\,eV$ is confirmed. It is easily to understand that the
closing of the Mott-gap and the opening of side Mott-gaps, so called replica, in the spectrum due to topological excitation are
classified as {\it non-trivial} topological effect.
In general the classification of correlated topological systems is a very
  active research field \cite{Rigol,Manmana,Rachel}. We refer to the section
  \ref{sec:FLO} in this article, where we show that in our theoretical
  results also at the crossing points the contribution by each Floquet mode
  is distinguishable and we analyse it as such.
For the investigation of the local density of states and the occupation
number, the filling, as well as the life-times of these states, an artificial cut-off of the Floquet
series, as is described in the literature, does not make sense from the numerical physics point of view of the
DMFT in frequency space, since it would hurt basically conservation laws. The
cut-off would lead to a drift of the overall energy of
the system, see section \ref{sec:FLO}. We refer thus to the bulk-boundary
correspondence \cite{KaneMele,HasanKane} which predicts that the results of
this work for bulk will be observed in a pump-probe experiment at the surface
of the semiconductor sample. In what follows we discuss des evolution of
  Floquet topological states for an increasing external driving frequency
  $\Omega_L$, see Fig.(\ref{IMEffectiveGFD}), and for an increasing amplitude of
  the driving, see Fig.(\ref{IMEffectiveGAD}), of the correlated system, and
  we discuss the physical consequences and possible applications.

\subsection{Topologically Induced Generation of Higher Harmonics and Optical Transparency}
\label{sec:HHG}

When we increase
the external excitation energy from $0\,eV$ to  $3.0\,eV$, we
find the evolution of Floquet topological quantum states, as well as the topologically induced Floquet band gaps for bulk matter. In Fig. \ref{IMEffectiveGFD}(a) the development of
a very clear Floquet fan for the valence as well as for the conduction band of the
correlated matter in the non-equilibrium is found.  Valence and
  conduction band split in a multitude of Floquet sub-bands which cross each
  other. For excitation energies up to $2.5\,eV$ the original band gap is
  subsequently closing, the first crossing point inside the semiconductor gap
  along the Fermi edge is found at $0.9\,eV$. With increasing excitation
  intensity, see Fig. \ref{IMEffectiveGFD}(b), higher order Floquet side bands
  emerge and we find the next prominent crossing point at the Fermi-edge for
  $0.45\,eV$. Band edges of higher order Floquet bands form crossing points
  with those of the first order, e.g. for the excitation energy  $0.7\,eV$
  the crossing points of the first side bands (02) with the higher number side
  bands are found at the atomic energy of $1.08\,eV$ and $2.3\,eV$ above the
  valence band edge and thus deep in the gap of the
  semiconductor. Semiconductors are well known for fundamental absorption at
  the band edge of the valence band. What we find here is that the absorption
  coefficient of the semiconductor is topologically modulated and non-trivial
  transitions at the crossing points of Floquet-valence subbands and
  Floquet-conduction subbands become possible and significant. Having in mind
  that a higher order Floquet subband is usually physically reached by absorption or
  generation  of higher harmonic procedures we find that there is a high
  probability existing for topologically induced direct transitions from the
  fundamental to higher order bands in those spectral positions where a Floquet band-edge crosses the
  inner band edge of the equilibrium valence band. At any bandedge in some
  sense directional scattering can be expected if the life-times of the
  states are on the scale of the expected scattering procedures.
In general the optical refractive index will be topologically modulated, electromagnetically induced
transparency may be observable for high amplitude excitations. The
topologically induced Floquet bands overlap and cross each other. Consequentially
very pronounced features such as narrow subgaps are formed
in the LDOS. In the excited semiconductor Floquet replica of valence and
conduction bands are formed and the dispersion is renormalized. We also find regions, e.g. for excitation
energies from $\hbar\Omega_L=0.9\,eV$ up to $\hbar\Omega_L=1.3\,eV$ and from
$\hbar\Omega_L=1.8\,eV$ up to $\hbar\Omega_L=2.2\,eV$ which show a topologically
induced metallic phase. The dressed states are the result of the Franz-Keldysh
effect \cite{Franz,Keldysh,Keldysh1,Keldysh2,Keldysh3} or AC-Stark effect which is well known for high intensity excitation
of semiconductor bulk and quantum wells \cite{SchmittRink,Chemla}. \\ 
From the methodological viewpoint of correlated electronics in the
non-equilibrium our results can be interpreted as follows. For finite
excitation frequencies an instantaneous transition to the topologically
induced Floquet band structure and a renormalized dispersion is derived. In the bulk system clear
Floquet bands develop, when the sample is excited by an intense
electrical field. This is observable in  Fig. \ref{IMEffectiveGFD}. \\
In Fig. \ref{IMEffectiveGAD} we show the same system as in  Fig. \ref{IMEffectiveGFD}
for constant driving energy of $1.75\,eV$ and an increasing driving
intensity up to $6.0\, MW/cm^2$. We find the evolution of side bands and an overall
vanishing semiconductor gap, which marks the transition from semi-conductor to a topologically
highly tunable and switchable conductor in the non-equilibrium.\\
At this point we do not investigate with our model the coupling to a geometrical edge or a resonator mode, which would lead for the optical case to additional contributions in
Eq. (\ref{Hamilton_we}) for the mode itself $\hbar\omega_o
a^{\dagger}a^{{\color{white}\dagger}}$  and the coupling of this resonator or
edge mode  $g\!\sum_{i, \sigma}
c^{\dagger}_{i,\sigma}c^{{\color{white}\dagger}}_{i,\sigma}(a^{\dagger}\! +
a)$.\\ $a^{\dagger}$ and $a$ are the creator and the annihilator of the
photon, $g$ is the coupling strength of the photonic mode to the electronic
system. From our results however we can conclude already that for
semi-conductor cavities and quantum wells as well as for structures which
enhance so called edge states that may couple to non-equilibrium topological
Floquet bands, these geometrical edge or surface resonances will induce an
additional topological effect within the full so far excitonic spectrum. It is
an additional effect beyond the bulk boundary correspondence. Dressed states may release energy quanta,
e.g. light, or an electronic current into the resonator mode \cite{PRB}. So we
expect from our results that such modes may become a very sensible switch in
non-equilibrium. \\
It is to clarify  whether novel topological effects in the non-equilibrium due to the geometry may
be of technological use, since they modify the full spectrum. For the investigation of ZnO as a laser material
in this regime the influences of surface resonators will be subject to
further investigations. It will be on target to find out all the signatures of
a topologically protected edge mode  in correlated and strongly correlated
systems out of equilibrium in the experiment and  to define the role of the
topologically induced local density
of bulk states e.g. for the occurrence of the electro-optical Kerr effect, the magneto-optical Kerr effect (MOKE) and the surface magneto-optical
Kerr effect (SMOKE). We believe that in correlated many-body
  systems out of equilibrium the topological edge mode is found to behave in the sense
of the bulk boundary correspondence. Those results are expected to become
modified or enhanced by a coupling of bulk states with a geometrical surface mode or a resonance.

\subsection{Floquet Sum and Consistency of the Numerical Framework}
\label{sec:FLO}

An analysis of the numerical validity in terms of the normalized and
frequency integrated density of states 

\begin{eqnarray}
N_{i}(\Omega_L)
:=
\int{\rm d}\omega N(\omega,\Omega_L)=1
\label{Eq:N_integrated}
\end{eqnarray}

\begin{figure}
\resizebox{0.5\textwidth}{!}{%
  \includegraphics{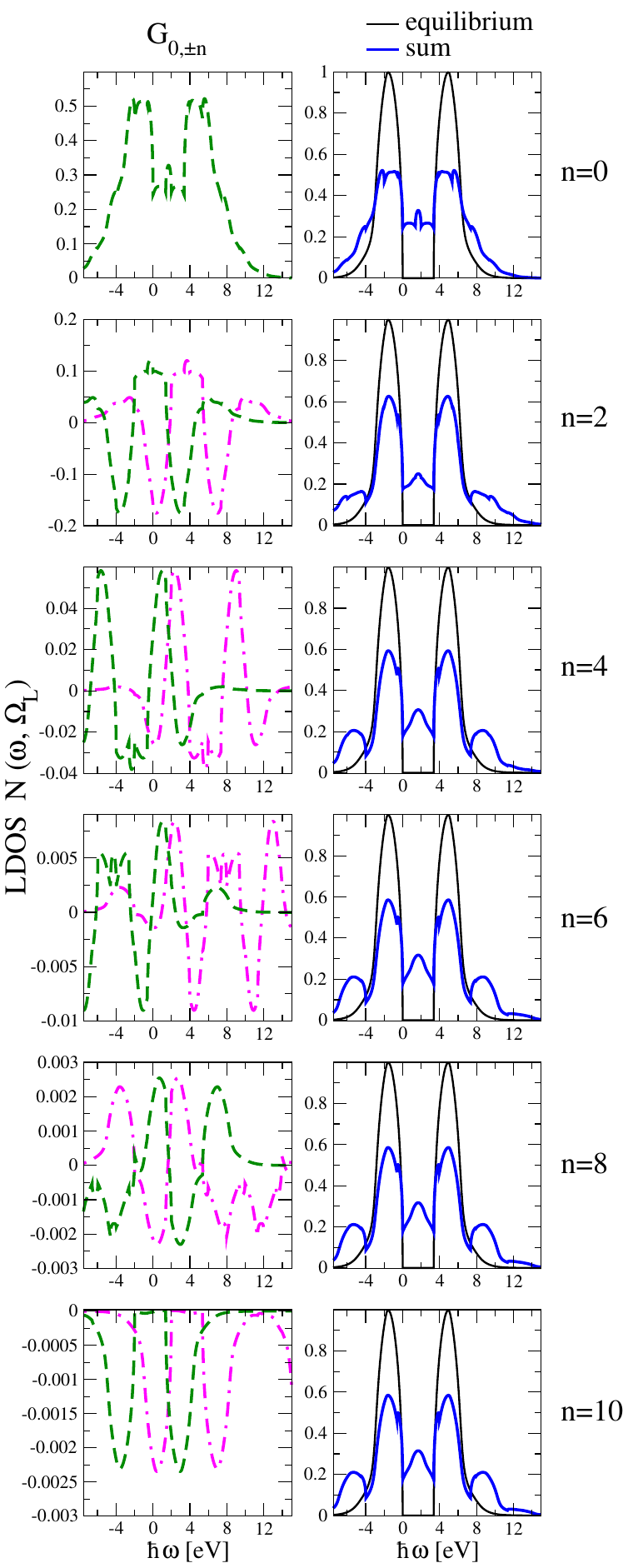}
}
\vspace{0.5cm}       
\caption{Floquet contributions and accuracy check of the numerical results for
  the LDOS of driven semiconductor bulk. The bandgap in equilibrium is $3.38 eV$, the Fermi edge is $1.69 eV$. For
  discussion see section \ref{sec:FLO}.}
\label{Floquet_Panel}       
\end{figure}

is generally confirmed for summing over Floquet indices up to the order or
10. We discuss in Fig. \ref{Floquet_Panel} on the l.h.s. the Floquet
contributions as examples with increasing order in steps of
$n=0,2,4,6,8,10$. With an increasing order of the Floquet index the amplitude of
the Floquet contribution decreases up to the point where the level of precision of
the DMFT self-consistency is reached. This is definitely the case for $n=10$
and thus a physical argument to cut the Floquet expansion off for $n=10$. The
displayed Floquet contributions $G_{0\pm n}$ are of course
perfectly mirror symmetric with respect to the Fermi edge, whereas the sum of
both contributions is directly symmetric with respect to the Fermi edge. This
is generally a proof of the validity of the numerical Fourier
transformation and the numerical scheme. The order of magnitude for Floquet
contributions of higher than  $n=4$ is almost falling consistently with the
rising Floquet index. We display here results for an external laser with the
wavelength of $\lambda=710.0 nm$ and the intensity of $3.8 MW/cm^2$. The
semiconductor gap is assumed to be $ 3.38 eV$ which corresponds to ZnO as a
laser active material. From the analysis it can be deduced that we include
Floquet contributions up to a precision of $10^{-3}$ with regard to their
effective difference from the final result which is displayed on the r.h.s of
Fig. \ref{Floquet_Panel} as the sum up to the order of n. That corresponds to
the overall accuracy of the self-consistent numerical approach. Higher order
modes do not contribute to the LDOS within the accuracy of the model under the
assumption of the given excitation intensities.\\
The physical interpretation of Fig. \ref{Floquet_Panel} is an interesting
point in the discussion of the evolution of the topological quantum
states. The order of the Floquet contribution $n$ numbers also the evolving
Floquet side bands which emerge in the LDOS, compare Fig. \ref{IMEffectiveGFD}(a). Whereas the lowest order Floquet contribution
$G_{00}$ is symmetric to the Fermi edge but strictly positive, higher order
contributions $G_{0\pm n}$ are mirror symmetric to each other and in sum they
can have negative contributions to the result of the LDOS. Their impact
decreases with the order. The Floquet contributions as such consequentially do
not have a direct physical interpretation however the sum of all contributions
equals the material characteristics of the local density of states, the LDOS.
Consequentially the increase of mathematical and
numerical precision of our framework of the Floquet-Keldysh DMFT by increasing
the number of Floquet contributions up to the level of the precision of the
DMFT has direct consequences for the finding and the accuracy of physical
results. Especially the investigation of the coupling of the driven electronic
system of the bulk with edge and surface modes will profit from this
precision. Since these coupling effects in nanostructure and waveguides are of
a high technological importance the advantage of the strictly controllable
precision of the 
Floquet DMFT approach in comparison to time dependent DMFT frameworks is quite
obvious.

\section{Conclusions}

We investigated in this article the evolution of Floquet topological quantum
states in wide band gap semiconductor bulk as a correlated
electronic system in the sense of a generalized Hubbard model with dynamical mean field
theory in the non-equilibrium. We found that non-equilibrium modulations
induce a non-trivial bandstructure which leads for several frequency ranges to
a topologically induced metal phase. This topological bandstructure is a
result of the AC-Stark effect. The intersection of Floquet bands and band-edges respectively induces
  transitions which enhance up- and downconversion effects as well as higher harmonic
generation. The absorption
  coefficient of the semiconductor is topologically modulated and non-trivial
  transitions at the crossing points of the equilibrium valence band, the
  Floquet-valence subbands, the Floquet-conduction subbands and the equilibrium conduction band become
  possible and their significance is depending on the excitation power. We also find the development of novel sub gaps as areas of
electromagnetically induced transparency.  The number of included Floquet
modes is chosen such that energy conservation is numerically guaranteed. This precision is essential in order to achieve
predictive results for semiconductor bulk which can be investigated for
optoelectronic and magneto-optoelectronic switching as well as  for innovative
laser systems.  Thus the bulk semiconductor under
non-equilibrium conditions as such has to be reclassified. It will be of great interest to investigate the interplay of the
here derived topological bulk effects with additional surface modifications
such as a surface resonance, a polariton coupling, or a surface
magneto-optical modulation.\\

{\bf Acknowledgments} The authors thank H. Monien, H. Wittel and F. Hasselbach for highly valuable discussions.

\section{Author contributions}
Both authors equally contributed to the presented work. Both authors were
equally involved in the preparation of the manuscript. Both authors have read
and approved the final manuscript.\\

%

\begin{thebibliography}{}
%

\bibitem{Kosterlitz} J. M. Kosterlitz, D. J. Thouless, {\em J. Phys. C : Solid
    State Phys.}, {\bf 6}, 1181 (1973).

\bibitem{KaneMele} L. Fu, C. L. Kane, E. J. Mele, {\em Phys. Rev. Lett.} {\bf
    98}, 106803 (2007). 

\bibitem{HasanKane} M. Z. Hasan, C. L. Kane, {\em Rev. Mod. Phys.} {\bf 82}, 3045 (2010).

\bibitem{Fu} L. Fu, C. L. Kane, {\em Phys. Rev. Lett.} {\bf 100}, 096407
(2008).

\bibitem{Moore}  G. Moore, N. Read, {\em Nucl. Phys. B} {\bf 360}, 362 (1991).

\bibitem{Dalibard} N. Goldmann, J. Dalibard, {\em Phys. Rev. X} {\bf 4}, 031027 (2014).


\bibitem{Zuti} I. Zuti, J. Fabian, S. Das Sarma, {\em Rev. Mod. Phys.}
{\bf 76}, 323 (2004).

\bibitem{Nayak} C. Nayak, S. H. Simon, A. Stern, M. Freedman, S.
Das Sarma, {\em Rev. Mod. Phys.} {\bf 80}, 1083 (2008).

\bibitem{Rechtsman} M. C. Rechtsman, J. M. Zeuner, Y. Plotnik, Y. Lumer,
  D. Podolsky, F. Dreisow, S. Nolte, M. Segev, A. Szameit {\em Nature} {\bf
    496}, 196 (2013). 

\bibitem{Demetrios} M. A. Bandres, S. Wittek, G. Harari, M. Parto, J. Ren,
  M. Segev, D. N. Christodoulides, M. Khajavikhan, {\em Science} {\bf 359}
  DOI:10.1126/science.aar4005 (2018).


\bibitem{Applsci2019} A. Lubatsch, R. Frank, {\em Appl. Sci.} {\bf 9(12)}, 2477 (2019).


\bibitem{Bernevig} B. A. Bernevig, T. L. Hughes, S.-C. Zhang, {\em Science}
{\bf 314}, 1757 (2006).

\bibitem{Knig} M. K\"onig, S. Wiedmann, C. Br\"une, A. Roth, H. Buhmann,
L. W. Molenkamp, X.-L. Qi, S.-C. Zhang, {\em Science}
{\bf 318}, 766 (2007).

\bibitem{Hsieh} D. Hsieh, D. Qian, L. Wray, Y. Xia, Y. S. Hor, R. J. Cava, M. Z. Hasan, {\em Nature} {\bf 452}, 970 (2008).

\bibitem{Xia} Y. Xia, D. Qian, D. Hsieh, L. Wray, A. Pal, H. Lin,
A. Bansil, D. Grauer, Y. S. Hor, R. J. Cava, M. Z.
Hasan, {\em Nature Physics} {\bf 5}, 398 (2009).

\bibitem{Zhang} H. Zhang, C.-X. Liu, X.-L. Qi, X. Dai, Z. Fang, S.-C.
Zhang, {\em Nature Physics} {\bf 5}, 438, (2009).

\bibitem{Lindner1} N. H. Lindner, G. Refael, V. Galitski, {\em Nature Physics} {\bf
    7} 490-495 (2011).


\bibitem{Podolsky} Y. Tenenbaum Katan, D. Podolsky,  {\em Phys. Rev. Lett.}
  {\bf 110}, 016802 (2013).


\bibitem{Kitagawa} T. Kitagawa, T. Oka, A. Brataas, L. Fu, E. Demler,
{\em Phys. Rev. B} {\bf 84}, 235108 (2011).

\bibitem{Gu} Z. Gu, H. A. Fertig, D. P. Arovas, A. Auerbach, {\em
    Phys. Rev. Lett.} {\bf 107}, 216601 (2011).



\bibitem{Gedik} Y. H. Wang, H. Steinberg, P. Jarillo-Herrero, N. Gedik {\em
    Science} {\bf 342}, 453 (2013).


\bibitem{Lindner2} M. S. Rudner, N. H. Lindner, E. Berg, M. Levin, {\em
    Phys. Rev. X} {\bf 3} 031005 (2013).

\bibitem{Lindner3} N. H. Lindner, D. L. Bergman, G. Refael, V. Galitski {\em
    Phys Rev. B} {\bf 87}, 235131 (2013).




\bibitem{Zoller} L. Jiang, T. Kitagawa, J. Alicea, A. R. Akhmerov, D.
Pekker, G. Refael, J. I. Cirac, E. Demler, M. D. Lukin, P. Zoller, {\em
  Phys. Rev. Lett.} {\bf 106}, 220402 (2011).


\bibitem{Rigol} L. D'Alessio, M. Rigol, {\em Nat. Comm.}, 6:8336 (2015).


\bibitem{Neupert} A. G. Grushin, A. Gomez-Leon, T. Neupert, {\em Phys. Rev. Lett.} {\bf 112}, 156801 (2014).

\bibitem{Bergholtz} E. Bergholtz, Z. Liu, {\em Int. J. Mod. Phys. B} {\bf 27},
  24, 1330017 (2013).


\bibitem{Floquet} G. Floquet, {\em Ann. l' Ecole Norm. Sup.} {\bf 12}, 47
(1883).

\bibitem{PRB} R. Frank, {\em Phys. Rev. B} {\bf 85}, 195463 (2012).

\bibitem{FrankANN} R. Frank, {\em Ann. Phys.} (Berlin) {\bf 525}, No. 1-2,
66-73 (2013).

\bibitem{Faisal} F. H. M. Faisal, J. Z. Kaminski, {\em Phys. Rev.  A} {\bf 56},
  1, 748 (1997).

\bibitem{Grifoni} M. Grifoni, P. H\"anggi, {\em Physics Reports} {\bf 304}, 5,
  229 (1998).

\bibitem{Brandes} S. Restrepo, J. Cerrillo, V. M. Bastidas, D. G. Angelakis,
  T. Brandes, {\em Phys. Rev. Lett.} {\bf 117}, 250401 (2016).

\bibitem{Eckardt} A. Eckardt, {\em Rev. Mod. Phys.} {\bf 89}, 011004-1 (2017).

\bibitem{Uhrig} M. H. Kalthoff, G.S. Uhrig, J. K. Freericks, {\em Phys. Rev. B} {\bf 98}, 035138 (2018).

\bibitem{Sentef} M. A. Sentef, M. Claassen, A. F. Kemper, B. Moritz, T. Oka, J. K. Freericks, T. P. Devereaux, {\em Nat. Comm.} {\bf 6}, 7047 (2015).

\bibitem{Fan} L. Yuan, S. Fan, {\em Phys. Rev. A} {\bf 92}, 053822 (2015).


\bibitem{ANN} A. Lubatsch, J. Kroha, {\em Ann. Phys.} (Berlin) {\bf 18}, No. 12, 863 - 867 (2009).

\bibitem{Kotliar} A. Georges, G. Kotliar, W. Krauth, M. J. Rozenberg, {\em Rev. Mod. Phys.} {\bf 68} (1), 13 (1996) 

\bibitem{Metzner} W. Metzner, D. Vollhardt, {\em Phys. Rev. Lett.} {\bf 62},
324 (1989)


\bibitem{Monien} P. Schmidt, H. Monien, arXiv:cond-mat/0202046 (2002).


\bibitem{Hettler} T. Maier, M. Jarrell, T. Pruschke, M. H. Hettler, {\em
    Rev. Mod. Phys.} {\bf 77}, 1027 (2005).

\bibitem{Freericks} J. K. Freericks, V. M. Turkowski, V. Zlatic, {\em Phys. Rev. Lett.} {\bf 97}, 266408 (2006).

\bibitem{OKA} N. Tsuji, T. Oka, H. Aoki, {\em Phys. Rev. Lett.} {\bf 103},
  047403 (2009).

\bibitem{Reichmann} N. Lin, C. A. Marianetti, A. J. Millis, D. R. Reichman
  {\em Phys. Rev. Lett.} {\bf 106}, 096402 (2011).

\bibitem{Zgid} D. Zgid, G. K.-L. Chan, {\em The Journal of Chemical Physics},
  {\bf 134} (9), 094115 (2011).



\bibitem{Werner} H. Aoki, N. Tsuji, M. Eckstein, M. Kollar, T. Oka, P. Werner, {\em Rev. Mod. Phys.} {\bf 86}, 779 (2014).

\bibitem{APLB} R. Frank, {\em Appl. Phys. B} {\bf 113}, 41 (2013).

\bibitem{NJP} R. Frank, {\em New J. Phys.} 15 123030 (2013)

\bibitem{Max} M. E. Sorantin, A. Dorda, K. Held, E. Arrigoni, {\em
    Phys. Rev. B} {\bf 97}, 115113 (2018).

\bibitem{Walter1} W. Hofstetter, T. Qin, {\em J. Phys. B: At. Mol. Opt. Phys.}
  {\bf 51} 082001 (2018).

\bibitem{Walter} T. Qin, W. Hofstetter, {\em Phys. Rev. B} {\bf 97}, 125115
  (2018).




\bibitem{SchmittRink} D. A. B. Miller, D. S. Chemla, T. C. Damen, A. C. Gossard, W. Wiegmann, T. H. Wood, C. A. Burrus, {\em Phys. Rev. Lett.} {\bf 53},
  22, 2173-2176 (1984).

\bibitem{Chemla} D. S. Chemla, W. H. Knox, D. A. B. Miller, S. Schmitt-Rink,
  J. B. Stark, R. Zimmermann, {\em Journal of Luminescence}
  {\bf 44}, 233-246 (1989).




\bibitem{Bukov} M. Bukov, L. D'Alessio, A. Polkovnikov, {\em Advances in
    Physics} {\bf 64}, No. 2, 139-226 (2015).


\bibitem{Zak89} J. Zak, {\em Phys. Rev. Lett.} {\bf 62}, 23, 2747-2750 (1989).

\bibitem{BerryElectronicProperties} D. Xiao, M.-C. Chang, Q. Niu, {\em
    Rev. Mod. Phys.} {\bf 82}, 3, 1959-2007 (2009). 


\bibitem{Vanderbilt2017} D. Gresch, G. Autes, O. V. Yazyev, M. Troyer,
  D. Vanderbilt, B. A. Bernevig, A. A. Soluyanov, {\em Phys. Rev. B} {\bf 95}, 075146 (2017).

\bibitem{JiaLu1} P.-C. Chang, J. G. Lu, {\em Appl. Phys. Lett.} {\bf 92},
  212113 (2008).

\bibitem{JiaLu2} P.-C. Chang, C.-J. Chien, D. Stichtenoth, C. Ronning, J. G. Lu,
{\em Appl. Phys. Lett.} {\bf 90}, 113101 (2007).


\bibitem{ZnOMono1} F. Huang, Z. Lin, W. Lin, J. Zhang, K. Ding, Y.  Wang,
  Q. Zheng, Z. Zhan, F. Yan, D. Chen, P. Lv,  X. Wang, {\em Chinese Science
    Bulletin} {\bf 59}, 1235 (2014).

\bibitem{ZnOMono2} W. I. Park, Y. H. Jun, S. W. Jung, G.-C. Yi, {\em
    Appl. Phys. Lett.} {\bf 82}, 6 964-966 (2003).



\bibitem{Rocksalt1} D. Fritsch, H. Schmidt, M. Grundmann, {\em
    Appl. Phys. Lett.} {\bf 88}, 134104 (2006).

\bibitem{Rocksalt2} H. Dixit, R. Saniz, D. Lamoen, B. Partoens, {\em
    J. Phys. Condens. Matter} {\bf 22}, 125505 (2010).

\bibitem{Wegener} M. Wegener, {\em Extreme Nonliner Optics}, ISBN 3-540-22291-X,
  Springer (2004).


\bibitem{Vanderbilt1} R. D. King-Smith and D. Vanderbilt, {\em Phys. Rev. B}
  {\bf 47}, 1651 (1993).

\bibitem{Vanderbilt2} D. Vanderbilt and R. D. King-Smith, {\em Phys. Rev. B}
  {\bf 48}, 4442 (1993).

\bibitem{Resta} R. Resta, {\em Rev. Mod. Phys.}, {\bf 66}, 3, 899-915 (1994).

\bibitem{DalLago} V. Dal Lago, M. Atala, L. E. F. Foa Torres, {\em
    Phys. Rev. A} {\bf 92}, 023624 (2015).

\bibitem{Gull} E. Gull, A. Millis, A. I. Lichtenstein, A. N. Rubtsov,
  M. Troyer, P. Werner, {\em Rev. Mod. Phys.}, {\bf 83}, 2, 349 (2011).


\bibitem{Hofstetter} M. Snoek, I. Titvinidze, C. Toke, K. Byczuk,
  W. Hofstetter, {\em New J. Phys.} {\bf 10}, 093008 (2008).


\bibitem{Peters1} R. Peters, T. Yoshida, H. Sakakibara, N. Kawakami, {\em
    Phys. Rev. B} {\bf 93}, 235159 (2016).

\bibitem{Peters2} R. Peters, T. Yoshida, N. Kawakami, {\em Phys. Rev.  B} {\bf
    98}, 075104 (2018).

\bibitem{Manmana} S. Manmana, A. M. Essin, R. M. Noack, V. Gurarie, {\em
   Phys. Rev. B} {\bf 86}, 205119 (2012).

\bibitem{Rachel} S. Rachel, {\em Rep. Prog. Phys.} {\bf 81}, 116501 (2018).

\bibitem{Franz} W. Franz, {\em Z. Naturforschung} {\bf 13a}, 484-489 (1958).

\bibitem{Keldysh} L. V. Keldysh, {\em J. Exptl. Theoret. Phys.} (USSR) {\bf
    33} 994-1003 (1957). 
\bibitem{Keldysh1}L. V. Keldysh, {\em Soviet Physics JETP} {\bf 6} 763-770 (1958).
\bibitem{Keldysh2} L. V. Keldysh, {\em J. Exptl. Theoret. Phys.} (USSR) {\bf
    47} 1945-1957 (1964).

\bibitem{Keldysh3} L. V. Keldysh,{\em Soviet
    Physics JETP} {\bf 20} 1307-1314 (1965).

\end{thebibliography}
%

\end{document}